\documentclass{article}
\usepackage{arxiv}
\usepackage[utf8]{inputenc}
\usepackage[T1]{fontenc}
\usepackage{hyperref}
\usepackage{url}
\usepackage{booktabs}
\usepackage{amsfonts}
\usepackage{nicefrac}
\usepackage{microtype}
\usepackage{graphicx}
\usepackage{amsmath}
\usepackage{multirow}
\usepackage{array}
\usepackage{longtable}
\usepackage{tabularx}
\usepackage{float}

\graphicspath{{./}}

\title{Robo-Reporters: Evaluating Autonomous AI Agents as Algorithmic Gatekeepers in Computational Journalism}

\author{
 Obada Kraishan \\
  College of Media and Communication\\
  Texas Tech University\\
  Lubbock, TX 79409, USA \\
 \And
 Kulsawasd Jitkajornwanich \\
   College of Media and Communication\\
  Texas Tech University\\
  Lubbock, TX 79409, USA \\
 \And
 Kerk Kee \\
  College of Media and Communication\\
  Texas Tech University\\
  Lubbock, TX 79409, USA \\
}

\begin{document}
\maketitle

\begin{abstract}
Artificial intelligence agents increasingly perform journalism tasks autonomously, searching for sources, evaluating credibility, and producing news content with minimal human oversight. Yet research has largely treated AI as a monolithic category, leaving the effects of architectural design unexamined. Drawing on gatekeeping theory, this study presents the first systematic comparison of four agent architectures, monolithic (Claude), chain-based (LangChain), multi-agent collaborative (CrewAI), and autonomous iterative (AutoGPT), across 200 controlled experiments spanning 50 journalism tasks of graduated difficulty. All architectures used the same underlying language model and identical tools, isolating architectural effects. Results revealed significant effects on task duration ($F(3, 196) = 24.54$, $p < .001$, $\eta^2 = .27$) and computational strategy ($F(3, 196) = 305.63$, $p < .001$, $\eta^2 = .82$), with architecture explaining 82\% of the variance in processing behavior. Multi-agent collaboration achieved the highest accuracy (84.7\%) at roughly twice the time cost of other designs. Multistage analysis of the monolithic architecture documented a 71.7\% source rejection rate, a quantitative parallel to classic human gatekeeping, while framework-based systems obscured their filtering inside abstraction layers. Transparency emerged as an architectural choice: framework designs excelled at structured attribution, whereas monolithic and iterative designs produced superior methodological documentation. Findings position architecture as a new structural level of gatekeeping and offer evidence-based guidance for newsrooms: chain-based designs for speed, multi-agent for accuracy, monolithic for versatility, and iterative for auditability.
\end{abstract}

\keywords{artificial intelligence \and autonomous agents \and computational journalism \and gatekeeping theory \and algorithmic transparency \and journalism automation}

\section{Introduction}

Over the past decade, artificial intelligence has changed journalism by growing from automating routine tasks to developing systems that independently make editorial decisions, such as selecting newsworthy stories and crafting narratives \cite{drr2016,carlson2015}. Contemporary AI agents can plan information-gathering strategies, dynamically evaluate sources, and iteratively refine outputs with minimal oversight \cite{beckett2019,diakopoulos2019,sumers2024,wang2024}. These capabilities position AI agents as genuine participants in the editorial process, fundamentally altering gatekeeping in contemporary journalism.

The architecture of an AI agent (its fundamental design pattern for processing information and making decisions) determines how an agent searches for information, which sources it consults, and eventually what information reaches audiences \cite{parasol2024,li2024,qian2024}. Despite the significance of architectural variations, mass communication research has generally overlooked them, resulting in newsrooms adopting technologies without empirical guidance on their editorial implications. This study addresses that gap through a controlled experimental comparison of four architectures representing major design paradigms: monolithic, chain-based, multi-agent collaborative, and autonomous iterative.

This gap has become urgent given the diversity of agent architectures now available to news organizations. Contemporary systems range from monolithic models that handle all reasoning internally to complex multi-agent configurations in which specialized components collaborate on information gathering and synthesis \cite{li2024,qian2024}. Some architectures prioritize speed and efficiency through streamlined processing pipelines, while others prioritize thoroughness through iterative refinement cycles. These choices embody different philosophies about how AI should participate in journalism: different architectures produce different journalistic outputs, just as different newsroom structures produce different editorial outcomes. Yet newsrooms currently adopt these systems with little empirical evidence about their consequences for core journalistic values such as accuracy, source diversity, and transparency \cite{broussard2019}. Because misunderstandings of what computational systems can and cannot do carry real institutional costs \cite{broussard2018}, systematic comparison of architectural alternatives has practical as well as theoretical significance.

Gatekeeping theory provides the central lens for this study \cite{shoemaker2009,singer2014}. While traditional gatekeeping research examined individual journalists and organizational structures \cite{white1950,breed1955}, contemporary scholarship has extended this framework to algorithmic systems \cite{napoli2014,thurman2019}. However, existing algorithmic gatekeeping work focuses primarily on recommendation and distribution systems rather than autonomous content-producing editorial agents. This study extends gatekeeping theory to autonomous agents by examining how architecture structures editorial decision-making at multiple production stages. Transparency requirements also warrant attention: architectural choices determine what decision processes become visible and what remains opaque \cite{diakopoulos2015,waddell2019}. This study investigates the following research questions:

\begin{itemize}
\item \textbf{RQ1:} How do different agent architectures compare in performance across journalism tasks of varying difficulty?
\item \textbf{RQ2:} How do gatekeeping patterns vary across agent architectures in terms of source selection, filtering, and diversity?
\item \textbf{RQ3:} How do transparency levels differ across agent architectures in documenting decision processes and methodological choices?
\end{itemize}

Given architectural diversity in design philosophy, we hypothesize systematic performance-transparency tradeoffs, where architectures optimizing for speed sacrifice thoroughness and vice versa (H1). We also expect multi-stage architectures to demonstrate higher gatekeeping selectivity than single-pass systems (H2).

By holding the underlying language model, tools, and tasks constant while varying only architecture, this study isolates architectural effects with a degree of experimental control rarely possible in field settings. In doing so, it offers three contributions: the first systematic comparison of agent architectures on journalism tasks, an extension of gatekeeping theory to autonomous editorial agents, and evidence-based guidance for newsrooms weighing adoption decisions.

\section{Literature Review}

\subsection{Computational Journalism and Agentic Architectures}

The integration of AI into journalism has advanced from template-based natural language generation for structured data \cite{thurman2017,clerwall2014} to systems capable of information synthesis, source credibility evaluation, and narrative construction without rigid templates \cite{diakopoulos2019,beckett2019,linden2017}. Contemporary AI agents powered by large language models can decompose high-level goals into actionable steps, select and apply tools dynamically, evaluate progress, and modify strategies based on intermediate results \cite{sumers2024,wang2024}; capabilities that position them as editorial participants rather than formatting tools. However, prior research typically treats AI as a monolithic category without examining how underlying system design influences outputs, leaving architectural effects confounded with task or model characteristics \cite{vanderkaa2014,jung2017}.

Audience research reinforces the importance of understanding these systems. Readers are generally unable to distinguish machine-generated from human-written articles when topics remain formulaic, though they judge automated content differently on dimensions such as credibility, expertise, and readability \cite{clerwall2014,graefe2018}, and perceived credibility shapes whether readers select automated journalism at all \cite{wlker2021}. The scholarly debate about what these capabilities mean for journalism remains unresolved. Some researchers situate AI journalism within longstanding trends toward the proceduralization and standardization of news work, viewing it as an acceleration of existing sociotechnical arrangements \cite{anderson2013,lewis2015}. Others contend that AI introduces genuinely novel forms of journalistic practice organized around different principles than human newswork \cite{deuze2019}. Either position, however, requires empirical understanding of how these systems actually behave, and that understanding remains thin at the level of system design.

The concept of autonomous agents itself has deep roots in artificial intelligence research on goal-directed systems capable of independent action \cite{russell2021}, but only the recent integration of large language models with agentic frameworks has made such autonomy practical for open-ended editorial work. Agent architecture, the design pattern governing how these capabilities integrate into a functional system, determines not just what an agent can do but how it approaches problems, allocates computational resources, and makes decisions under uncertainty.

Major architectural paradigms differ in their approach to editorial tasks \cite{li2024,parasol2024}. Monolithic architectures process tasks through unified workflows within a single model instance, prioritizing cognitive coherence but potentially sacrificing modularity and transparency \cite{anthropic2024}. Chain-based architectures decompose tasks into sequential stages, underlining predictability through clear separation of concerns but risking error propagation across steps \cite{chase2022,chase2024}. Multi-agent architectures distribute work across specialized agents (mirroring human newsroom organization) enabling high performance through specialization while introducing coordination overhead \cite{qian2024,li2024}. Autonomous iterative architectures enable independent goal decomposition and self-refinement through multiple cycles, prioritizing thoroughness over speed \cite{significantgravitas2024}. Existing benchmarks comparing these architectures focus mainly on technical domains such as coding and mathematical reasoning, leaving journalism-specific evaluation criteria like source diversity and transparency unstudied.

\subsection{Gatekeeping Theory and Algorithmic Transparency}

Gatekeeping theory traces how information is selected, filtered, and transmitted from sources to audiences \cite{shoemaker2009}. White's \cite{white1950} foundational study of a wire editor established how personal preferences and practical constraints shape published stories; ensuing work expanded this to organizational routines, professional norms, economic pressures, and social forces across five levels of analysis \cite{breed1955,gans1979,shoemaker1991}. Algorithmic gatekeeping research has documented how automated systems in news distribution and recommendation introduce systematic biases (preferential treatment of engagement-optimizing content, underrepresentation of certain perspectives) through encoded rules and learned patterns \cite{napoli2014,thurman2019,bakshy2015,gillespie2014}. These systems raise explicitly normative questions about what kinds of gatekeepers machines ought to be, including concerns about filter bubbles and the fragmentation of shared information environments \cite{nechushtai2019}. However, distribution-focused research tells us little about how AI systems perform editorial gatekeeping when producing content. Diakopoulos \cite{diakopoulos2019} and Carlson \cite{carlson2015} propose that AI journalism involves gatekeeping decisions across source selection, evaluation, extraction, synthesis, and presentation stages---each requiring editorial judgment previously made by humans. Singer \cite{singer2014} argues this algorithmic basis for decisions prompts reevaluation of how journalistic values are encoded in automated systems.

Transparency has emerged as a central concern in algorithmic journalism, serving functions of audience credibility assessment, professional oversight, and public trust maintenance \cite{karlsson2011,diakopoulos2015,waddell2019}. Diakopoulos \cite{diakopoulos2015} distinguishes input, process, and output transparency as distinct dimensions requiring different technical implementations. Research on explainable AI has developed visualization and feature importance methods, but these often prove insufficient for complex multistep reasoning involving planning and iterative refinement \cite{ribeiro2016,arrieta2020}. Importantly, transparency requirements may conflict with architectural choices that optimize for speed or accuracy---a tradeoff requiring empirical investigation. Accountability frameworks for AI journalism identify both forward-looking responsibility and backward-looking answerability as essential, with architectural choices determining how readily decisions can be traced and systems modified \cite{diakopoulos2018,waddell2019}. This study contributes by measuring transparency empirically across architectural approaches and examining its relationships with performance and gatekeeping dimensions.

Taken together, three gaps emerge from this literature. First, research on AI journalism has evaluated systems in isolation or treated AI as a single category, leaving architectural effects unexamined. Second, algorithmic gatekeeping scholarship has concentrated on distribution and recommendation, providing little evidence about gatekeeping within autonomous content production. Third, transparency research has proposed normative frameworks but rarely measured how design choices shape what becomes visible in practice. The present study addresses these gaps through a controlled experimental comparison of four agent architectures completing identical journalism tasks with the same underlying model and tools. This design isolates architecture as the explanatory variable across three dimensions that prior work has treated separately: performance, gatekeeping behavior, and transparency. In doing so, the study connects the computational literature on agent design with mass communication scholarship on editorial control, testing whether gatekeeping theory extends to a class of systems its founders could not have anticipated.

\section{Methodology}

\subsection{Agent Implementations}

\begin{figure}[tbp]
\centering
\includegraphics[width=0.98\textwidth]{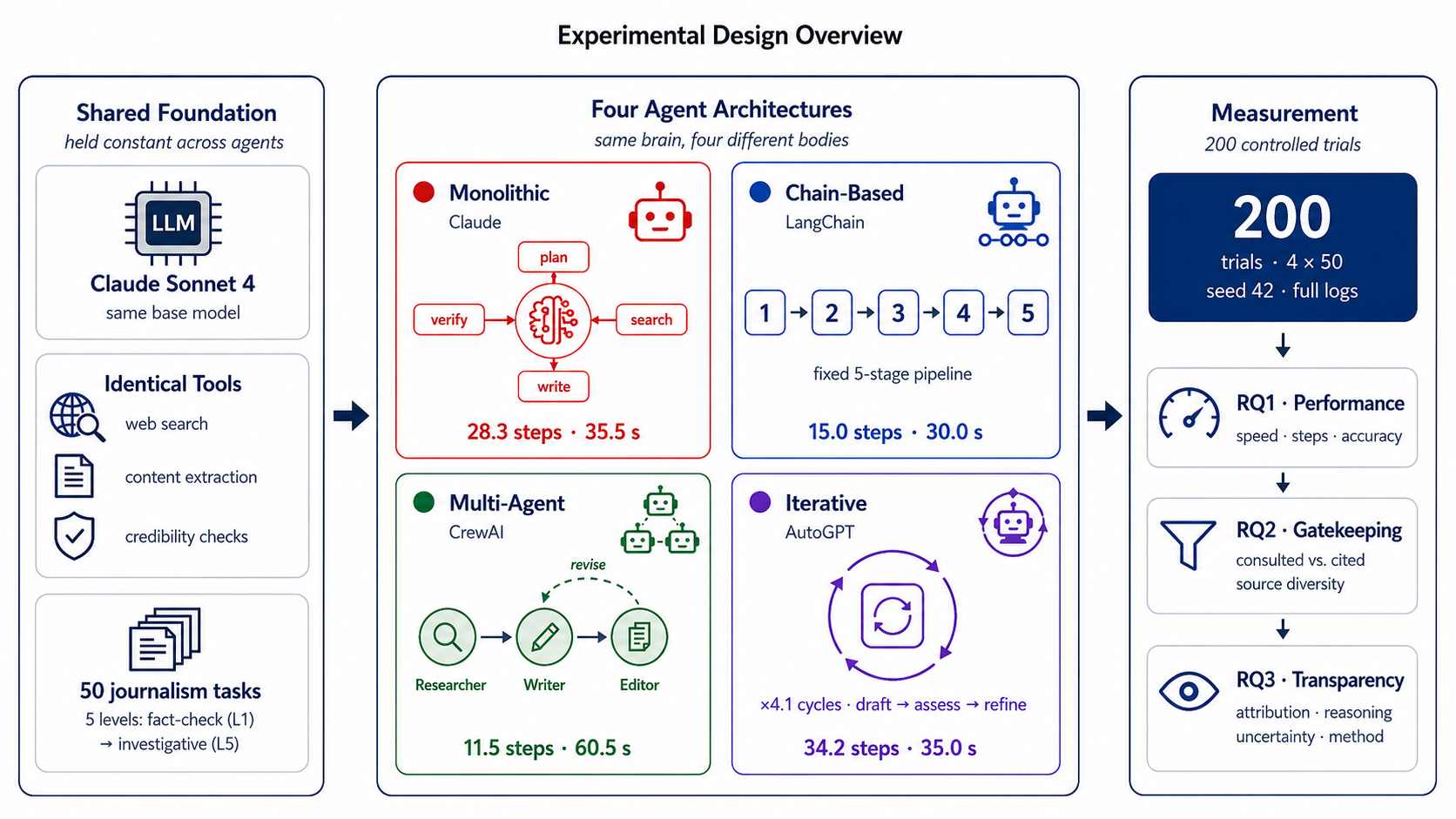}
\caption{Experimental design overview. Four agent architectures, all running the same base model with identical tools and an identical 50-task battery, completed 200 controlled trials evaluated through three measurement lenses corresponding to the research questions.}
\label{fig:design}
\end{figure}

We implemented four distinct agent architectures, all using Claude Sonnet 4 (claude-sonnet-4) as the underlying language model to isolate architectural effects from model capability differences. Each agent had access to identical tools: web search (via Tavily and DuckDuckGo APIs), web content extraction (BeautifulSoup and Playwright), and source credibility assessment. This controlled design ensures observed differences reflect architectural choices rather than varying capabilities or tool availability (see Figure~\ref{fig:design}).

The monolithic agent (Claude) processes journalism tasks through integrated workflows within a single model instance, generating a comprehensive plan and executing searches and synthesis through continuous reasoning without explicit stage boundaries. Implementation included custom prompt engineering for journalistic reasoning, explicit sourcing instructions, and comprehensive logging of all tool calls. The agent averaged 28.3 reasoning steps per task.

The chain-based agent (LangChain, v0.1.0) decomposes tasks into five sequential stages: task analysis, information gathering, source evaluation, synthesis, and output formatting. Each stage receives output from the previous stage, highlighting modularity and predictability. The chain-based agent averaged 15.0 steps per task with zero variance, reflecting its fixed pipeline.

The multi-agent collaborative system (CrewAI, v0.1.0) distributes tasks across three specialized agents: researcher, writer, and editor. The researcher gathers information, the writer constructs narratives, and the editor reviews for accuracy and completeness, with the ability to request additional information or revisions. This division mirrors human newsroom organization. The system averaged 11.5 steps per task.

The autonomous iterative agent (AutoGPT, v0.5.0) enables independent goal decomposition and iterative refinement, executing an autonomous loop until self-assessment indicates satisfactory output. Iteration count and specific actions vary based on task complexity rather than following predetermined sequences. The agent averaged 4.1 iterations per task with 34.2 total steps.

\subsection{Journalism Task Design}

We developed 50 journalism tasks spanning five difficulty levels based on journalism education curricula and professional practice standards \cite{filak2019,deuze2005}. All tasks involved information gathering from public web sources, source evaluation, and synthesis with proper attribution. Level 1 ($n = 10$): Fact verification tasks requiring confirmation of specific claims using reliable sources. Level 2 ($n = 10$): Event coverage tasks requiring comprehensive summaries of recent occurrences. Level 3 ($n = 10$): Timeline construction tasks tracing developments across multiple sources over time. Level 4 ($n = 10$): Source analysis tasks evaluating information credibility and comparing coverage across outlets. Level 5 ($n = 10$): Investigative research tasks requiring discovery of non-obvious connections, conflicts of interest, and complex entity relationships. Ground truth answers were developed through extensive author research and verification, providing benchmarks while acknowledging journalism's inherent judgment calls.

\subsection{Experimental Procedures}

We conducted 200 controlled experiments (4 agents $\times$ 50 tasks $\times$ 1 repetition) in January 2026. Task assignment used block randomization with seed value 42, ensuring each agent received identical tasks while preventing order effects. All experiments ran on identical hardware with consistent internet connectivity. Each trial proceeded with no human intervention during agent execution, and all agents completed tasks within timeout limits (5 minutes for simple, 10 minutes for complex tasks). Comprehensive logging captured all tool calls with parameters and results, all reasoning steps and intermediate decisions, execution time with millisecond precision, token usage, source consultation and citation decisions, and any errors or retry attempts.

\subsection{Evaluation Metrics}

Performance metrics included task duration (elapsed time in seconds), computational steps (distinct operations per task), success rate (task completion without errors), and accuracy (comparison to ground truth via automated similarity metrics and keyword matching). For fact verification tasks, binary accuracy assessed yes/no correctness; for other tasks, accuracy reflected key fact completeness.

Gatekeeping metrics tracked sources consulted (all sources accessed), sources cited (explicitly attributed in outputs), citation rate (cited/consulted ratio), source diversity (unique domain count), and Gini coefficient (source distribution inequality). Where architecture allowed, we conducted multistage analysis across consultation, evaluation, selection, and presentation phases.

Transparency metrics operationalized Diakopoulos's \cite{diakopoulos2015} framework across four weighted dimensions: source attribution (0.30), reasoning visibility (0.25), uncertainty acknowledgment (0.20), and methodological transparency (0.25). Composite transparency scores were calculated as:

\begin{equation}
T_{overall} = 0.30\,T_{attribution} + 0.25\,T_{reasoning} + 0.20\,T_{uncertainty} + 0.25\,T_{methodological}
\end{equation}

where each $T_{dimension}$ ranges from 0 (completely opaque) to 1 (fully transparent). Weights reflect relative importance for journalistic accountability, with attribution weighted most heavily for enabling verification and oversight.

Statistical analysis used one-way ANOVA to examine architectural effects on performance and transparency, with eta-squared effect sizes and Bonferroni-corrected post-hoc pairwise comparisons. Chi-square tests examined categorical gatekeeping features. All tests used $\alpha = .05$.

\section{Results}

All 200 experimental trials completed successfully, providing complete data for analysis and demonstrating robust autonomous operation across all four architectures. The following sections present results for each research question in turn: performance differences (RQ1), gatekeeping patterns (RQ2), and transparency (RQ3), using one-way ANOVAs with Bonferroni-corrected pairwise comparisons throughout.

\subsection{RQ1: Performance Differences Across Agent Architectures}

Task duration. LangChain demonstrated the fastest average completion ($M = 30.0$ s, $SD = 20.2$), followed by AutoGPT ($M = 35.0$, $SD = 23.7$) and Claude ($M = 35.5$, $SD = 12.9$). CrewAI required substantially longer ($M = 60.5$, $SD = 20.0$), nearly double the fastest architecture. Table~\ref{tab:duration} presents full descriptive statistics.

\begin{table}[htbp]
\centering
\caption{Descriptive Statistics for Task Duration by Agent Architecture}
\label{tab:duration}
\begin{tabular*}{\textwidth}{@{\extracolsep{\fill}}lrrrrrr}
\toprule
Architecture & $M$ & $SD$ & Median & Min & Max & $n$ \\
\midrule
LangChain & 30.01 & 20.18 & 25.14 & 14.86 & 119.29 & 50 \\
AutoGPT   & 34.95 & 23.66 & 27.88 & 14.88 & 129.45 & 50 \\
Claude    & 35.51 & 12.91 & 32.89 & 22.49 & 99.19  & 50 \\
CrewAI    & 60.49 & 20.03 & 55.79 & 39.88 & 130.04 & 50 \\
\bottomrule
\end{tabular*}

\vspace{2pt}
\parbox{\textwidth}{\footnotesize \textit{Note.} Duration measured in seconds. $M$ = mean; $SD$ = standard deviation.}
\end{table}

\begin{figure}[tbp]
\centering
\includegraphics[width=0.85\textwidth]{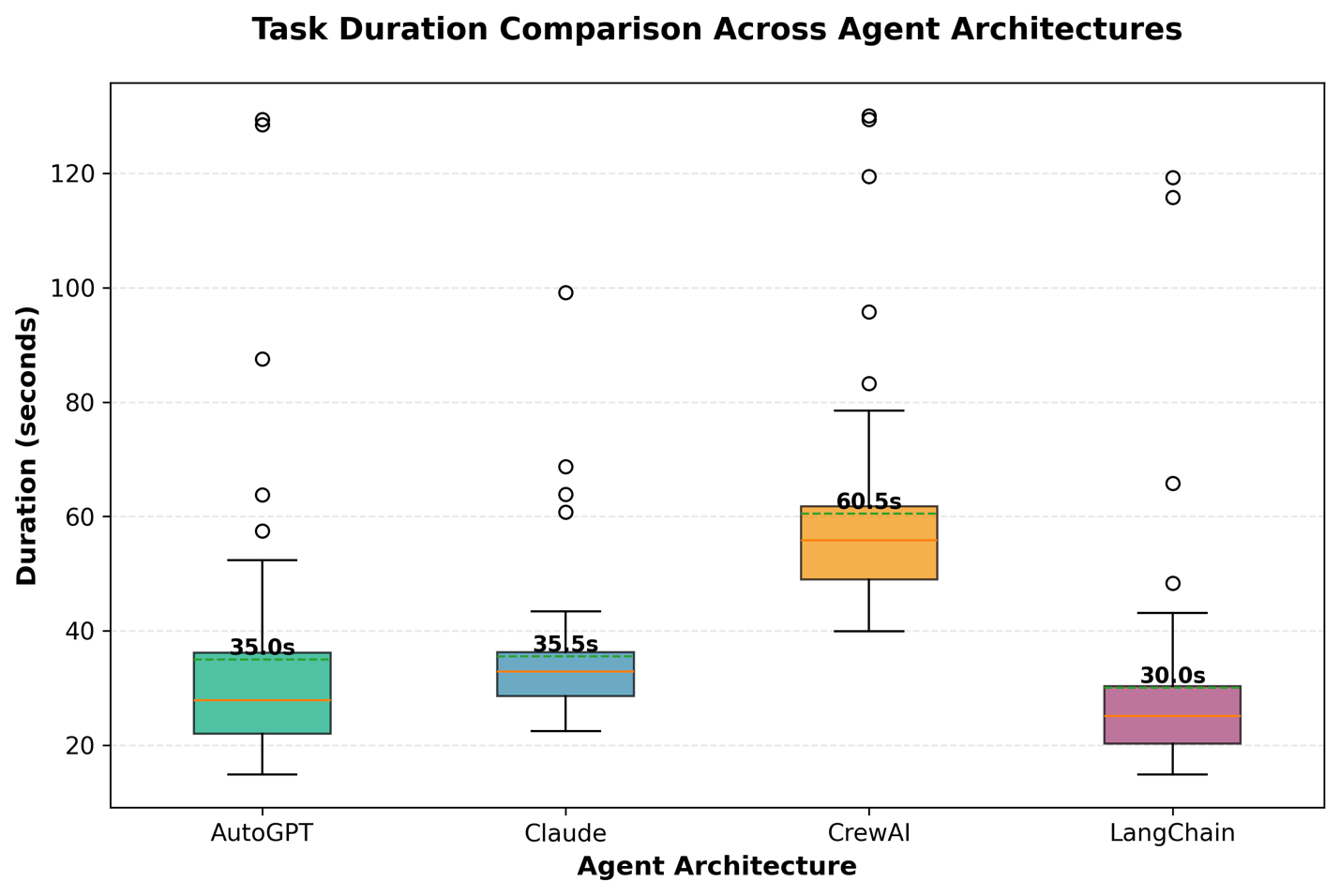}
\caption{Task duration distributions across five difficulty levels by agent architecture. \textit{Note.} Box plots show median (center line), interquartile range (box), and whiskers extending to 1.5$\times$ IQR. Architectural differences increase with task complexity.}
\label{fig:1}
\end{figure}

One-way ANOVA confirmed these differences reached statistical significance, $F(3, 196)$ = 24.54, $p < .001$, $\eta^2 = .27$, indicating architecture accounted for 27.3\% of variance in task completion time; a large effect by conventional standards \cite{cohen1988}. Post-hoc Bonferroni comparisons showed CrewAI required significantly more time than LangChain ($MD = 30.5$ s, $p < .001$, $d = 1.52$), AutoGPT ($MD = 25.5$ s, $p < .001$, $d = 1.17$), and Claude ($MD = 25.0$ s, $p < .001$, $d = 1.48$). All three comparisons showed large effect sizes (d $>$ .80). The three faster architectures did not differ significantly from each other (all $p$s $>$ .10). Architectural differences magnified with task complexity: for Level 5 investigative tasks, CrewAI's median duration exceeded 80 seconds while LangChain maintained sub-40 second medians (see Figure~\ref{fig:1}).

\begin{figure}[tbp]
\centering
\includegraphics[width=0.72\textwidth]{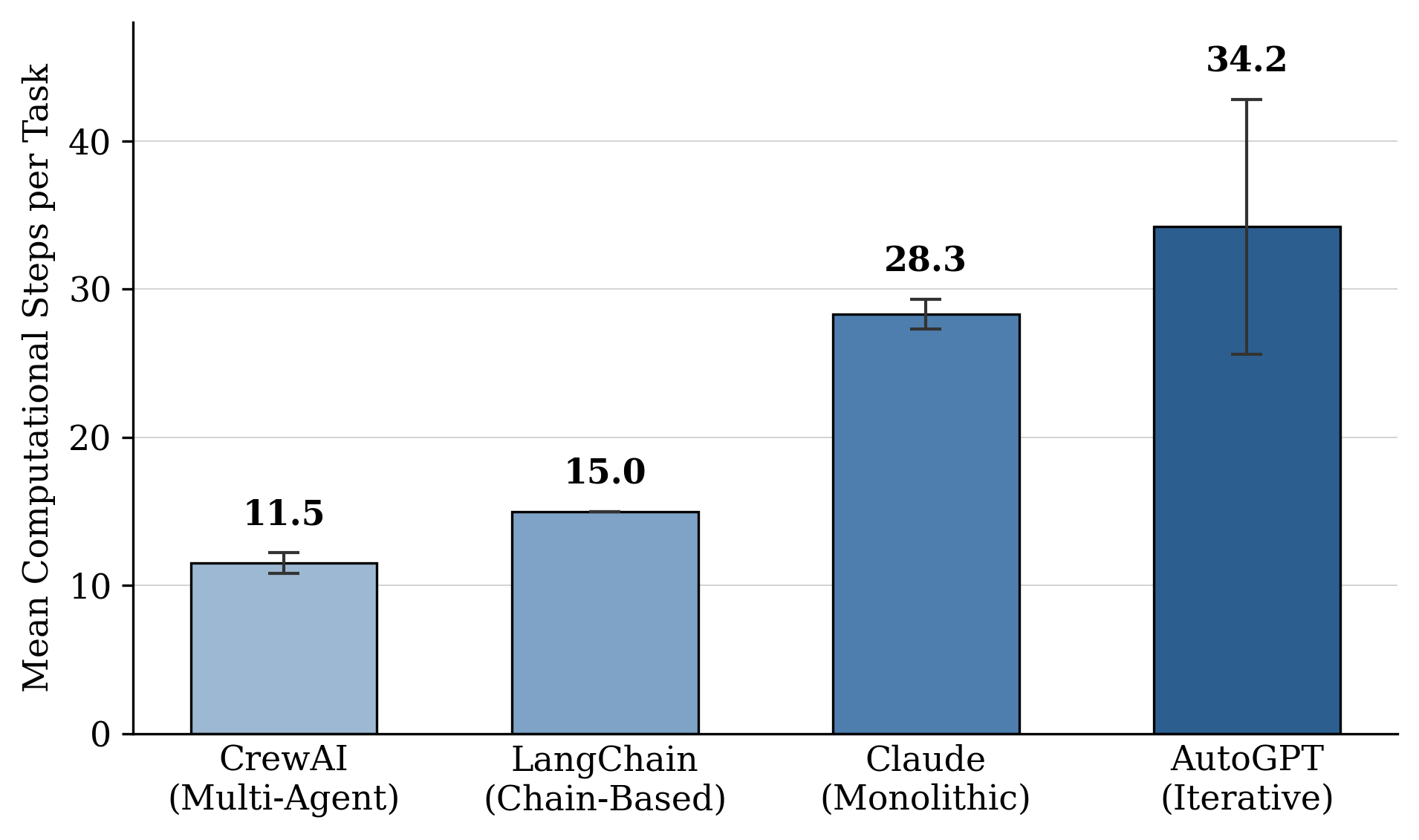}
\caption{Mean computational steps required by each agent architecture. \textit{Note.} Error bars represent standard deviations. CrewAI's collaborative specialization achieves high efficiency through coordinated agent roles, while AutoGPT's iterative refinement produces the highest and most variable step counts.}
\label{fig:2}
\end{figure}

Computational efficiency. Analysis of processing steps revealed even more pronounced architectural effects. CrewAI completed tasks in an average of 11.5 steps ($SD = 0.7$), substantially fewer than other architectures. LangChain required exactly 15.0 steps ($SD = 0.0$) per task. Claude averaged 28.3 steps ($SD = 1.0$). AutoGPT showed the highest step counts ($M = 34.2$, $SD = 8.6$), consistent with its iterative refinement approach. One-way ANOVA revealed extraordinarily strong effects, $F(3, 196)$ = 305.63, $p < .001$, $\eta^2 = .82$, indicating architecture accounted for 82.4\% of variance---an exceptionally large effect rarely observed in behavioral research. All pairwise comparisons reached significance after Bonferroni correction (all $p$s $<$ .001), with effect sizes ranging from $d = 0.96$ (AutoGPT vs. Claude) to $d = 19.16$ (Claude vs. CrewAI; see Figure~\ref{fig:2}).

Output accuracy. CrewAI achieved the highest mean accuracy (84.7\%, $SD = 18.2$), with 35 of 50 tasks (70\%) receiving perfect scores. LangChain demonstrated 76.9\% accuracy ($SD = 22.1$, 29 perfect scores), Claude 75.5\% ($SD = 23.4$, 28 perfect scores), and AutoGPT 70.6\% ($SD = 25.8$, 27 perfect scores). Despite these differences, one-way ANOVA did not reveal statistically significant architectural effects on accuracy, $F(3, 196)$ = 2.43, $p = .066$, $\eta^2 = .04$. Accuracy patterns across difficulty levels showed all architectures exceeding 85\% on Level 1 tasks, with CrewAI maintaining higher accuracy on investigative Level 5 tasks (78\%) compared to AutoGPT's lowest accuracy on the same items (64\%). For source analysis tasks (Level 4), CrewAI's specialized editor agent produced accuracy exceeding other architectures by approximately 12 percentage points. All four architectures achieved 100\% task success rates across all 200 trials (Table~\ref{tab:accuracy}).

\begin{table}[htbp]
\centering
\caption{Accuracy and Source Citation Patterns by Agent Architecture}
\label{tab:accuracy}
\begin{tabular*}{\textwidth}{@{\extracolsep{\fill}}lrrrr}
\toprule
Architecture & Mean Accuracy & Perfect Scores & Sources Cited ($M$) & Unique Domains \\
\midrule
CrewAI    & 84.7\% & 35/50 & 2.5 & 94 \\
LangChain & 76.9\% & 29/50 & 3.0 & 104 \\
Claude    & 75.5\% & 28/50 & 4.2 & 139 \\
AutoGPT   & 70.6\% & 27/50 & 3.5 & 117 \\
\bottomrule
\end{tabular*}

\vspace{2pt}
\parbox{\textwidth}{\footnotesize \textit{Note.} Accuracy represents mean similarity to ground truth answers. Perfect scores indicate exact matches.}
\end{table}

\subsection{RQ2: Gatekeeping Patterns Across Agent Architectures}

\begin{figure}[tbp]
\centering
\includegraphics[width=0.85\textwidth]{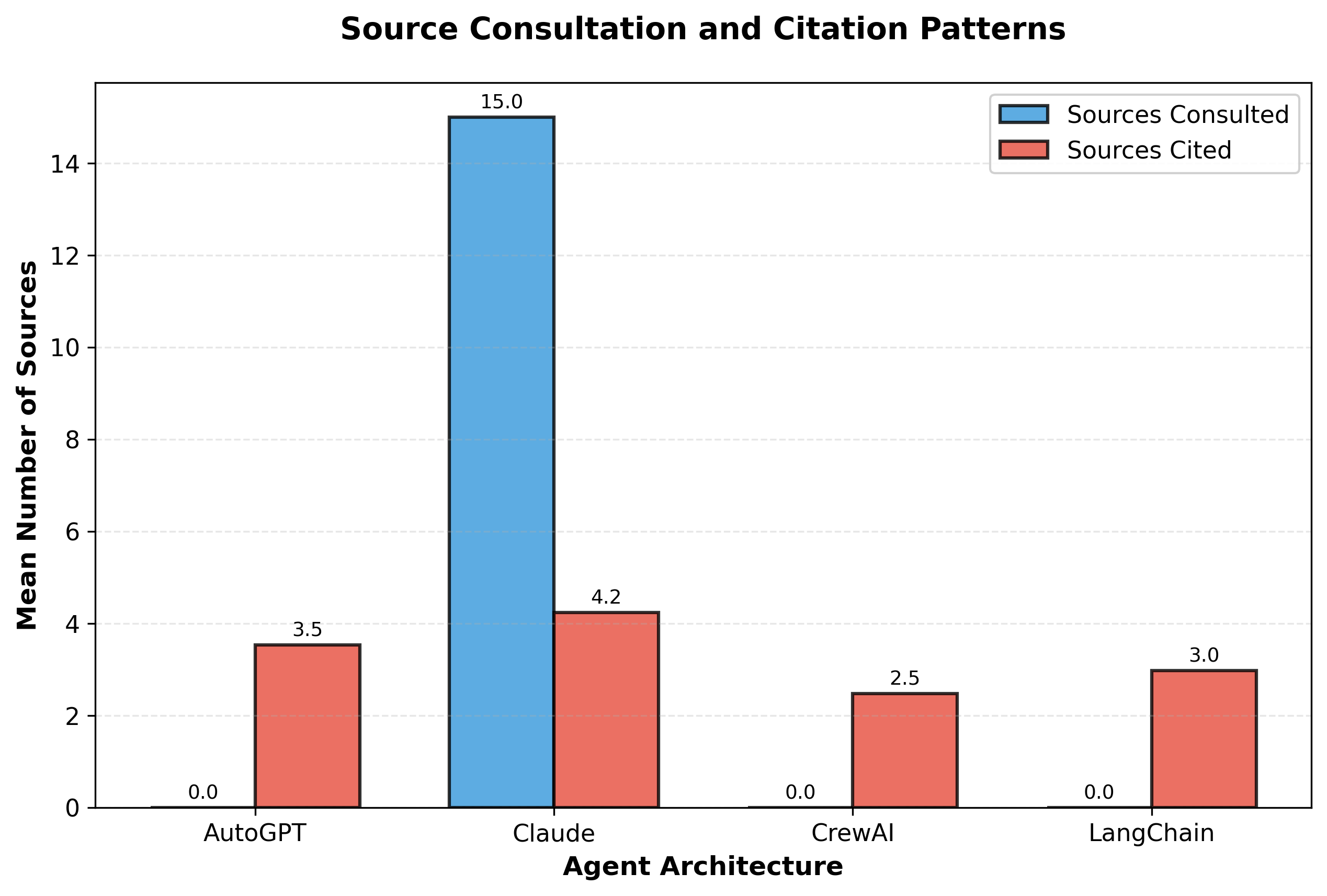}
\caption{Source consultation and citation patterns. \textit{Note.} Gap between consultation and citation bars indicates gatekeeping selectivity. Only the Claude architecture enabled consultation tracking.}
\label{fig:3}
\end{figure}

Source selection and filtering. Examination of source consultation and citation patterns uncovered a fundamental architectural difference in gatekeeping visibility. Claude's monolithic architecture enabled complete tracking of source consultation: across all tasks, Claude consulted an average of 15.0 sources ($SD = 8.2$) through web searches, ultimately citing 4.2 ($SD = 1.8$) in final outputs---a citation rate of 28.3\% and rejection rate of 71.7\% ($SD = 12.4$). In contrast, LangChain, CrewAI, and AutoGPT utilized framework-based web search implementations that handled source retrieval internally without explicit logging, enabling only final citation counts (3.0, 2.5, and 3.5 respectively). This measurement limitation itself constitutes a meaningful finding: monolithic implementations enable complete process visibility, while framework-based approaches prioritize efficiency at the cost of gatekeeping transparency (see Figure~\ref{fig:3}).

Among architectures, CrewAI's multi-agent design produced the most conservative citation patterns ($M = 2.5$, $SD = 1.2$), reflecting the editor agent's strict quality standards. AutoGPT cited slightly more ($M = 3.5$, $SD = 1.6$), with iterative refinement enabling discovery of additional sources. LangChain cited 3.0 sources on average ($SD = 1.4$), and Claude cited the most ($M = 4.2$, $SD = 1.8$). For Claude's multistage analysis, the largest filtering occurred between consultation and evaluation (54.7\% of consulted sources did not receive full content extraction), with lower attrition at evaluation-to-selection (38.2\%). High-authority sources showed higher selection rates (64\% cited after consultation) compared to general news (42\%) or personal blogs (18\%); recent sources were cited more than older ones (55\% vs. 38\%).

Source diversity. Claude accessed the most diverse source base (139 unique domains across 50 tasks), followed by AutoGPT (117), LangChain (104), and CrewAI (94). Table~\ref{tab:diversity} presents complete diversity metrics. All architectures cited Wikipedia as their most-used domain, though dependence varied: CrewAI showed highest concentration (20.2\% of citations) while Claude showed lowest (10.4\%). Gini coefficient analysis indicated relatively equal distribution in CrewAI (.237) and moderate concentration in Claude (.300). Domain type coding revealed all architectures drew primarily from news outlets (45-52\%) and Wikipedia/reference sources (18-28\%), with smaller proportions from academic (8-12\%), government (10-15\%), and other sources---no significant architectural differences in domain type distribution emerged, $\chi^2(12) = 18.3$, $p = .11$.

\begin{table}[htbp]
\centering
\caption{Source Diversity Metrics by Agent Architecture}
\label{tab:diversity}
\begin{tabular*}{\textwidth}{@{\extracolsep{\fill}}lrlrrr}
\toprule
Architecture & Unique Domains & Top Domain & Top Domain \% & Diversity Score & Gini Coefficient \\
\midrule
Claude    & 139 & Wikipedia & 10.4\% & 0.90 & .300 \\
AutoGPT   & 117 & Wikipedia & 20.3\% & 0.80 & .319 \\
LangChain & 104 & Wikipedia & 16.1\% & 0.84 & .281 \\
CrewAI    & 94  & Wikipedia & 20.2\% & 0.80 & .237 \\
\bottomrule
\end{tabular*}

\vspace{2pt}
\parbox{\textwidth}{\footnotesize \textit{Note.} Diversity score = 1 $-$ top domain concentration. Gini coefficient ranges from 0 (perfect equality) to 1 (perfect inequality).}
\end{table}

\subsection{RQ3: Transparency Across Agent Architectures}

Overall transparency scores. AutoGPT achieved the highest composite transparency ($M = .491$, $SD = .124$), followed closely by Claude ($M = .486$, $SD = .094$) and LangChain ($M = .481$, $SD = .107$). CrewAI demonstrated the lowest transparency ($M = .450$, $SD = .101$). One-way ANOVA revealed significant but modest architectural effects, $F(3, 196)$ = 2.89, $p = .036$, $\eta^2 = .04$, indicating measurable differences accounting for limited overall variance relative to within-architecture variation.

Transparency dimensions. Source attribution revealed the largest architectural differences: LangChain and CrewAI achieved the highest attribution scores ($M = .831$ and .825), substantially exceeding AutoGPT (.531) and Claude (.340). These differences primarily reflected structured source list inclusion---LangChain provided explicit source sections in 74\% of outputs and CrewAI in 76\%, compared to AutoGPT's 32\% and Claude's 2\%. Claude and AutoGPT instead employed inline citations at higher rates (4.2 and 3.5 per output), demonstrating narrative integration versus structured separation philosophies.

Reasoning visibility remained relatively low across all architectures (Claude: $M = .282$; LangChain: .200; CrewAI: .176; AutoGPT: .148), with most outputs including methodological statements (90-100\%) but rarely explaining specific decisions. Uncertainty acknowledgment was highest for Claude ($M = .388$; uncertainty expressions in 64\% of outputs) followed by CrewAI (.312, 52\%), LangChain (.284, 48\%), and AutoGPT (.272, 44\%). Explicit inability statements appeared in only 0-8\% of outputs across architectures. Methodological transparency showed the most pronounced architectural difference: AutoGPT achieved the highest score ($M = .962$, $SD = .041$) reflecting detailed autonomous logs, with Claude close behind (.945). LangChain (.500) and CrewAI (.383) scored substantially lower due to framework abstraction obscuring internal operations. Average step counts correlated strongly with methodological transparency (\textit{r} = .78, $p < .001$). Additional transparency features---separation of facts from interpretation (22-42\%), acknowledgment of alternative perspectives (2-12\%), and disclosure of limitations (6-18\%)---showed universal limitations across all architectures.

\section{Discussion}

\subsection{Theoretical Implications}

Our findings extend gatekeeping theory by identifying architecture as a structural level of analysis between organizational choice and technical implementation. Just as newsroom organizational structure constrains which stories receive prominence, agent architecture constrains which sources receive consultation and which information reaches audiences. Claude's documented multistage gatekeeping dynamics (broad initial consultation of 15 sources, narrower evaluation of 6.8 sources, and selective final citation of 4.2 sources with a 71.7\% rejection rate) parallel White's \cite{white1950} "Mr. Gates" funnel patterns in quantitative, algorithmic form. This advances theoretical understanding by demonstrating that architectural design instantiates multistage editorial judgment comparable to human decision hierarchies. Architectural gatekeeping differs from human gatekeeping, however: human gatekeepers apply tacit professional judgment enabling flexible responses to unique circumstances \cite{breed1955,gans1979}, while algorithmic gatekeepers produce more consistent but less nuanced decisions through explicit rules and learned patterns.

The transparency findings complicate Diakopoulos's \cite{diakopoulos2015} claim that algorithmic journalism requires new transparency forms due to opacity. Different architectures make different decision aspects visible: monolithic architectures enable comprehensive operation logging (superior methodological transparency), while framework-based approaches excel at structured output attribution. Transparency is not unitary but multidimensional, requiring conscious design choices across attribution, reasoning, uncertainty, and methodological dimensions. Performance-transparency tradeoffs also proved more complex than simple efficiency-quality dichotomies. CrewAI achieved highest accuracy while requiring most time but fewest steps, demonstrating that collaborative specialization can simultaneously optimize multiple objectives---challenging assumptions that automation necessarily sacrifices quality for speed \cite{singer2014,carlson2015}.

\subsection{Performance Tradeoffs and Practical Implications}

The performance results revealed architectural specializations rather than universal rankings. LangChain\'s chain-based approach optimized for speed (30 s average) through streamlined sequential processing, suitable for breaking news contexts with straightforward requirements. CrewAI\'s collaborative design prioritized accuracy (84.7\%) despite requiring 60 seconds, fitting investigative journalism where thoroughness matters more than speed. The editor agent\'s quality control instantiated newsroom organizational practices in algorithmic form. Claude\'s monolithic approach balanced speed, accuracy, and transparency across all metrics without excelling at any single dimension, suited for versatile general-assignment newsrooms. AutoGPT\'s autonomous iterative design emphasized thoroughness through self-refinement and achieved the highest transparency scores, fitting investigative contexts prioritizing quality assurance and process documentation.

These specializations suggest newsrooms should align architectural choices with editorial priorities rather than seeking universally optimal solutions. The gatekeeping findings underscore the importance of architectural transparency during procurement: newsrooms should require vendors to document how systems consult sources, evaluate credibility, and make citation decisions. Framework-based architectures offer straightforward deployment but at transparency costs, as frameworks obscure internal operations. Monolithic and autonomous architectures require more custom implementation but enable comprehensive instrumentation. The universal limitations in uncertainty acknowledgment and alternative perspective presentation, given that all architectures showed explicit inability statements in only 0-8\% of outputs, suggest systematic challenges beyond architectural choice, possibly stemming from training objectives rewarding definitive answers.

\subsection{Limitations}

Several limitations qualify these findings. First, we examined only four architectures from the broader design space; hybrid architectures combining elements may achieve different tradeoff patterns. Second, all architectures employed the same underlying language model (Claude Sonnet 4), limiting generalizability to other model families; architectural effects may interact with model capabilities. Third, tasks represented only a subset of journalism work, primarily information gathering and synthesis amenable to automated evaluation, excluding interviewing, cultivating confidential sources, and other human-centric functions. Fourth, automated accuracy evaluation via keyword matching and semantic similarity did not assess narrative quality, stylistic appropriateness, or audience engagement; human expert evaluation would provide richer quality assessment. Fifth, we examined agents in isolated experimental contexts rather than integrated newsroom environments, where infrastructure constraints and workflow integration may produce different outcomes. Sixth, transparency measurement challenges with framework-based architectures limited our ability to calculate rejection rates or multistage filtering patterns for LangChain, CrewAI, and AutoGPT, a limitation that itself reflects a meaningful finding about framework opacity.

\subsection{Future Research}

Future research should replicate these architectural comparisons across different base models (e.g., GPT-4, Gemini, Llama) to determine whether the observed patterns reflect general architectural properties or model-specific interactions. Task diversity should expand toward journalism activities involving human sources, extended investigations, and ethical decision-making, and expert journalist evaluations should complement automated accuracy metrics with assessments of narrative quality. Hybrid architectures combining chain-based speed with multi-agent quality control merit systematic testing, as do methods for instrumenting framework-based systems so that their internal gatekeeping becomes observable. Finally, field studies of deployed systems in working newsrooms would test whether the laboratory tradeoffs documented here persist under production conditions.

\section{Conclusion}

This study provides the first systematic comparison of autonomous AI agent architectures in journalism contexts, representing how fundamental design choices shape editorial performance, gatekeeping patterns, and transparency. Our controlled experimental comparison across 50 journalism tasks demonstrated substantial architectural effects on task completion speed ($\eta^2 = .27$) and computational efficiency ($\eta^2 = .82$), with more modest effects on accuracy and transparency. These findings extend gatekeeping theory by revealing architecture as a structural level determining information flows from sources to audiences, with Claude's 71.7\% rejection rate documenting substantial algorithmic selectivity comparable to human gatekeeping funnels. Performance results challenge simplistic automation narratives by showing that different architectures optimize for different objectives: chain-based approaches for speed (30 s average), collaborative approaches for accuracy (84.7\%), and iterative approaches for multidimensional balance. Transparency findings reveal that architectural design powerfully shapes which decision aspects become visible---framework-based architectures excelling at structured source attribution, monolithic and autonomous approaches providing superior methodological documentation. Universal limitations in uncertainty acknowledgment suggest systematic challenges in how contemporary AI systems handle epistemic humility, warranting attention in both technical design and regulatory frameworks. These contributions inform theory by demonstrating architecture as a gatekeeping variable, and practice by providing evidence-based guidance for newsrooms matching architectural approaches to editorial contexts. As AI systems assume greater editorial autonomy, thoughtful architectural selection based on transparent evaluation is essential for maintaining journalistic quality, accountability, and public trust.

**\
**

\end{document}